\documentclass{article}
\usepackage{spconf,amsmath,graphicx}
\usepackage{booktabs, multirow} 
\usepackage{amssymb} 
\usepackage{hyperref} 
\title{PAMA-TTS: Progression-Aware Monotonic Attention for Stable Seq2Seq TTS With Accurate Phoneme Duration Control}
\name{Yunchao He \qquad Jian Luan \qquad Yujun Wang}
\address{Xiaomi Corporation, Beijing, China}
\begin{document}
\topmargin=0mm 
%
\maketitle
\begin{abstract}
Sequence expansion between encoder and decoder is a critical challenge in sequence-to-sequence TTS. Attention-based methods achieve great naturalness but suffer from unstable issues like missing and repeating phonemes, not to mention accurate duration control. Duration-informed methods, on the contrary, seem to easily adjust phoneme duration but show obvious degradation in speech naturalness. This paper proposes PAMA-TTS to address the problem. It takes the advantage of both flexible attention and explicit duration models. Based on the monotonic attention mechanism, PAMA-TTS also leverages token duration and relative position of a frame, especially countdown information, i.e. in how many future frames the present phoneme will end. They help the attention to move forward along the token sequence in a soft but reliable control. Experimental results prove that PAMA-TTS achieves the highest naturalness, while has on-par or even better duration controllability than the duration-informed model.


\end{abstract}
\begin{keywords}
Alignment guidance, duration control, attention mechanism, seq2seq TTS, speech synthesis
\end{keywords}
\section{Introduction}
\label{sec:intro}
Text-To-Speech is a typical sequence-to-sequence modeling task. In general, its input is a grapheme or phoneme sequence while the output is a much longer sequence of acoustic parameters at the frame level. In recent popular encoder-decoder architectures, the attention mechanism demonstrated strong capability in mapping two sequences with different lengths \cite{Attention} and achieved high naturalness in TTS tasks \cite{Tacotron,Tacotron2,TransformerTTS}. However, for unseen texts, it may also bring errors like missing and repeating phonemes, unexpected long silence, and even failure to produce speech completely \cite{FF-attention, SM-attention, LR-attention}. Many efforts have been made to enhance the attention robustness by constraining the attention to meet locality, monotonicity, and completeness, such as Forward attention \cite{FF-attention}, Stepwise monotonic attention \cite{SM-attention}, and Location-Relative attentions \cite{LR-attention}. However, none of them constrained how many frames one token should occupy. Without it, phonemes in an unseen text may still be articulated extremely short or too long in synthesized speech.

Other than attention-based methods, many studies utilize a separate duration model to implement the sequence upsampling. Fastspeech \cite{FastSpeech}, Fastspeech2 \cite{FastSpeech2}, and DurIAN \cite{DurIAN} duplicate encoder outputs according to the phoneme duration. Non-Attentive Tacotron \cite{Non-Atten-Tacotron} implements upsampling with Gaussian weights. The ground-truth duration is obtained from external forced-alignment tools \cite{FastSpeech2, DurIAN, Non-Atten-Tacotron, P-Tacotron, Tacotron2-duration} or by internal joint training \cite{ALIGNTTS, JDI-T}. Regardless of how the alignment is obtained and how the duplicated tokens are smoothed, duration-informed methods always show naturalness degradation due to hard duration control.


Differentiable duration models \cite{e2e-adv, P-Tacotron2} are also designed. They need no phoneme alignment guidance but to optimize duration model parameters by minimizing the final spectrogram reconstruction loss directly. For the duration loss, only the total duration of phonemes in a sequence is taken into account. It improved the naturalness of duration-informed methods. However, in such networks, the output of the duration model may not physically stand for phoneme duration. Particularly, when the predicted duration of one word is adjusted when inference, the durations of other words in the synthesized speech are often affected unexpectedly. 


This paper proposes a Proceeding-Aware Monotonic Attention (PAMA\footnote[1]{Audio examples: \url{https://pama-tts.github.io/}} ) for sequence-to-sequence TTS to realize accurate phoneme duration control without naturalness degradation. The neural network is based on Tacotron2 but the Location Sensitive Attention (LSA) is replaced by stepwise monotonic attention \cite{SM-attention}. Besides, a soft guidance attention matrix is generated from ground-truth alignment to benefit both the efficiency of attention training and the correctness of learned alignment. At the same time, an auxiliary duration model is trained with the same alignment label. From the duration model, latent duration representation and backward position embedding are offered to attention memory and query respectively. The main contributions of this paper include:
\begin{itemize}
    \item Design an innovative guidance attention matrix for alignment constraint. The guidance is soft at phoneme boundaries since there are no solid ground-truth breaks;
    \item Introduce latent duration representation into encoder output as attention memory. With this information, alignment loss converges faster and more stably;
    \item Introduce backward frame position within phoneme into prenet output as an attention query. In this way, the generation of current spectrum conditions on not only the preceding spectrogram but also how many future frames the present phoneme should end within. The former ensures the spectrum smoothness while the latter helps the phoneme duration control. Their impacts are balanced by the network dynamically.
\end{itemize}


\section{Related Works}
\label{sec:rela}

Although PAMA-TTS calculates attention alignment vector recursively in the same way as stepwise monotonic attention in \cite{SM-attention}, both attention query and memory of them are different. For query, PAMA-TTS adds backward position information for token proceeding awareness. For memory, PAMA-TTS adds latent duration representation for efficient and stable training convergence.

VAENAR-TTS \cite{VAENAR} introduces a latent variable \(Z\) to help soft attention alignment, in which \(Z\) implicitly stands for phoneme duration. However, there are no phoneme level duration labels to guide \(Z\) explicitly. Besides, VAENAR-TTS leverages both annealing reduction factor and causality mask to help attention-based alignment learning other than applies monotonic constraint.

Moreover, the attention alignment loss in PAMA-TTS is quite similar to PAG in \cite{Guided-Attention}. However, PAMA-TTS generates guidance matrices in a softer way for better flexibility, since the results of a forced alignment tool may have slight distortion, especially on found data.

\section{Architecture}
\label{sec:arch}

\begin{figure}[htb]
  \centering
  \includegraphics[scale=1.0]{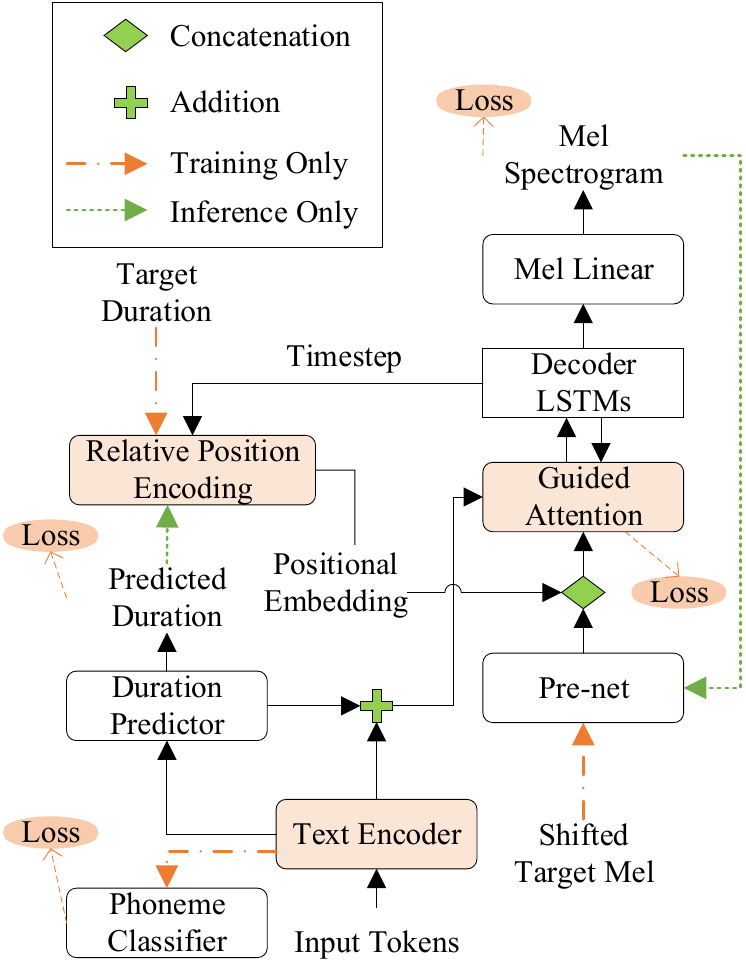}
  \caption{Architecture diagram of PAMA-TTS. The yellow and green dotted lines are turned on only for the training and inference stage respectively. }
  \label{fig:icassp_2022_fig2}
\end{figure}

The architecture of PAMA-TTS is shown in Fig. ~\ref{fig:icassp_2022_fig2}. Tacotron2 \cite{Tacotron2} with stepwise monotonic attention \cite{SM-attention} is employed as the backbone. Modified modules are highlighted and will be illustrated below one by one.




\subsection{Text Encoder \& Phoneme Classifier}
\label{ssec:encoder}

\begin{figure}[htb]
  \centering
  \includegraphics[scale=1.0]{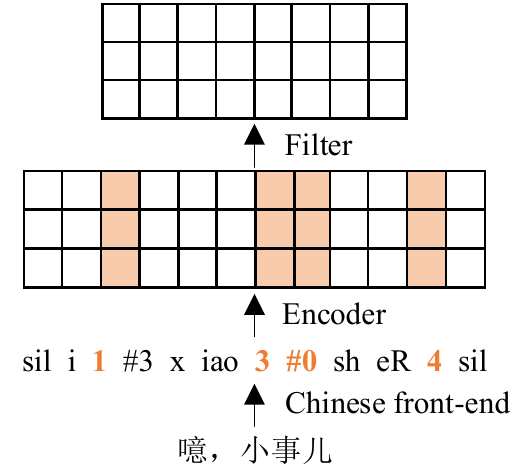}
  \caption{An illustration of how to skip some hidden states (highlighted in yellow) of encoder output. The symbols \#0, \#1, \#2, and \#3 denote the boundaries of syllables, prosodic words, intermediate phrases, and intonational phrases respectively. The numbers 1-5 denote tones of the previous syllable. }
  \label{fig:icassp_2022_fig1}
\end{figure}

The text encoder takes a sequence of token IDs as inputs and outputs the latent representation of them, which consist of regular phonemes, tones, prosodic boundaries, and silence. The tokens are placed in a carefully designed order to build up input sequences as demonstrated in Fig. ~\ref{fig:icassp_2022_fig1}. 

Since tones and most prosodic boundaries (except intonation phrase boundary \#3, which can be regarded as silence or short pause as well) do not correspond to any acoustic frames in speech, a filter is applied to skip the hidden states of them as shown in Fig. ~\ref{fig:icassp_2022_fig1}. A similar strategy is used in DurIAN \cite{DurIAN}, but they remove only prosodic boundaries. Moreover, the trimmed encoder output is fed into a phoneme classifier to ensure the token location information remains. Both above designs aim at making the subsequent alignment learned by the attention mechanism more meaningful.

The encoder structure is the same as that of Tacotron2, i.e. three convolutional layers followed by a BLSTM layer. For the phoneme classifier, a single feed-forward layer with softmax cross-entropy loss is employed.

\subsection{Guided Attention Matrix }
\label{ssec:guided}

\begin{figure}[htb]
\begin{minipage}[b]{0.48\linewidth}
  \centering
  \centerline{\includegraphics[width=4.0cm]{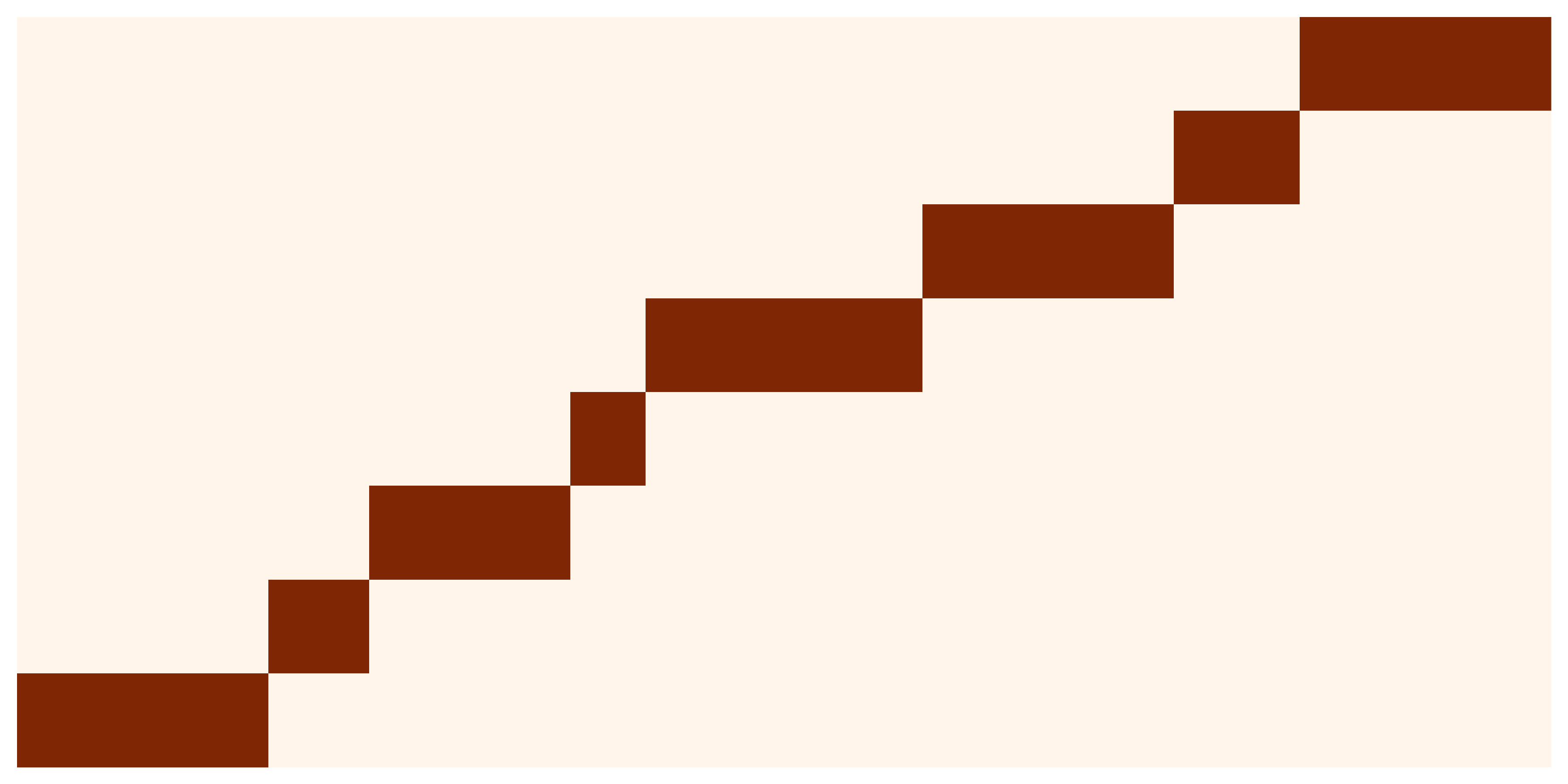}}
  \centerline{(a) Hard guidance matrix}\medskip
\end{minipage}
\hfill
\begin{minipage}[b]{0.48\linewidth}
  \centering
  \centerline{\includegraphics[width=4.0cm]{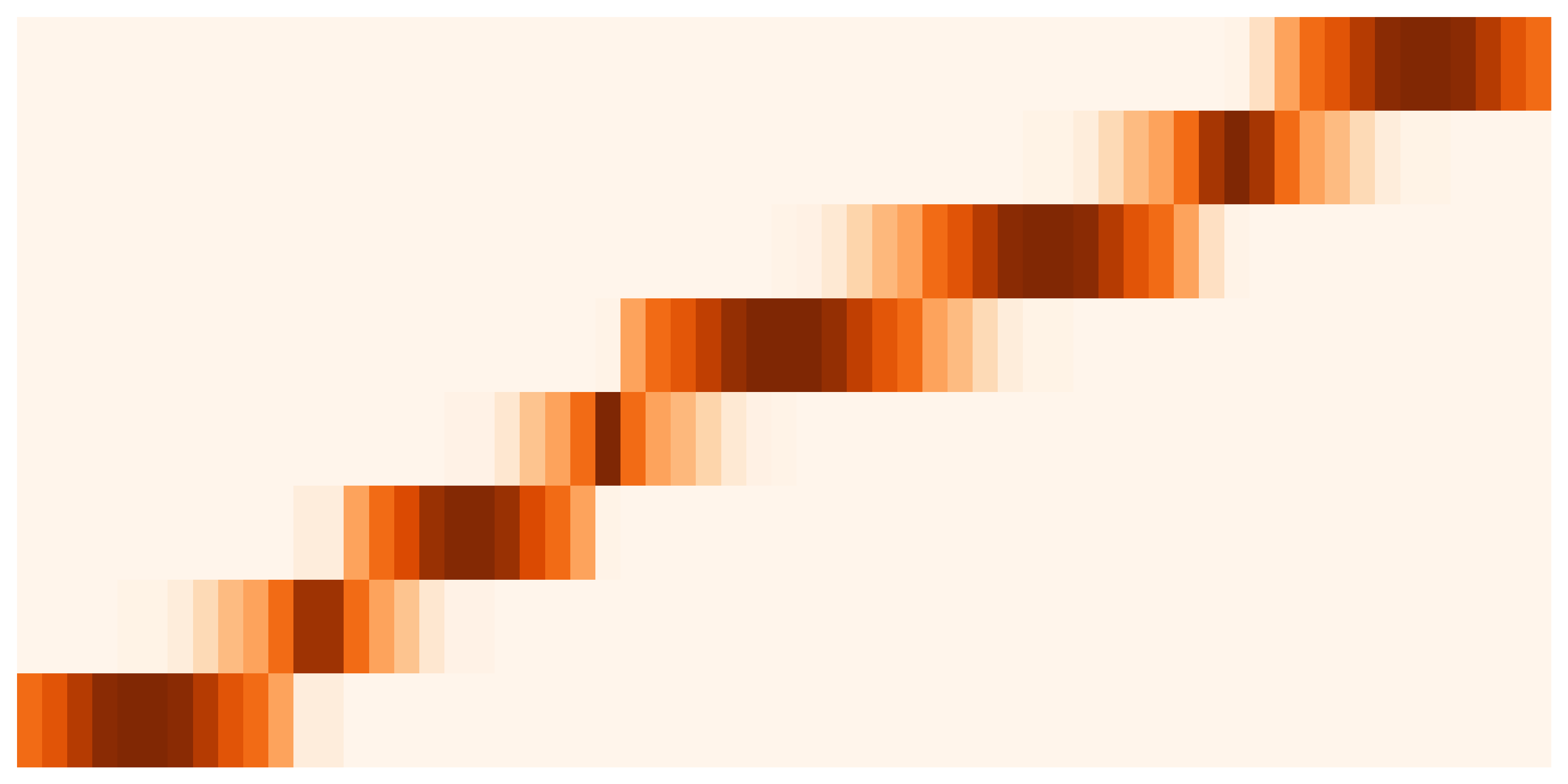}}
  \centerline{(b) Fuzzy guidance matrix}\medskip
  \label{fig:fuzzy_matrix}
\end{minipage}
\caption{An example of hard and fuzzy alignment guidance. Weights in (b) change smoothly at phoneme boundaries.}
\label{fig:hard_vs_fuzzy}
\end{figure}

Guided attention is used to help the attention module learn a correct mapping between phoneme sequence and acoustic frames efficiently. Previous work \cite{Guided-Attention} used time-aligned phoneme sequences obtained by forced alignment to generate hard guidance matrices. Considering the existence of alignment errors, this paper improves the guidance matrix to have fuzzy weights at phoneme boundaries as shown in Fig. ~\ref{fig:hard_vs_fuzzy}.

According to statistics on large data, most alignment errors of phonemes are within 3 frames. Therefore, the weights at boundaries in the guidance matrix are linearly transitioned from 0 to 1 in six frames with a step size of 0.2. Then, a mean square error is computed as alignment loss as
\[ L_{align} = \frac{1}{T} \sum_{i=1}^{T} \sum_{j=1}^{N} (W _{ij} - \alpha _{ij})^2 \tag{1} \]
where \(T\), \(N\) denote the number of spectrogram frames and filtered tokens, \(W,\alpha  \in \mathbb{R}^{N \times T} \) are the guidance matrix and attention weight matrix, respectively. 



\subsection{Progression-Aware Monotonic Attention}
\label{ssec:pama}


The proposed PAMA is based on stepwise monotonic attention. To make the monotonic attention aware of the mapping progression between phonemes and spectrogram, two additional pieces of information is leveraged: one is a latent duration code for attention memory, and the other is a relative position embedding for attention query. 

The latent duration code is from the last hidden layer of a duration predictor and transformed by a linear layer. For each phoneme, its duration code is added with its encoder output to generate key and value for the attention mechanism. In this way, the attention's key and value vectors carry duration information more explicitly. 

The relative position embedding is a concatenation of two vectors from learnable look-up tables. One is for the forward position within a phoneme, which implies the distance to the beginning of the token. The other is for the backward position, which denotes the distance to the end of the token. Both of the two distances are ceilinged with a constant \( C \). For each acoustic frame, its relative positional embedding is concatenated with the output of prenet to generate an attention query. 




Generally speaking, prenet output only carries information of the preceding spectrogram. The injection of relative position embedding, especially bringing the knowledge that how many future frames the current phoneme should end within, helps the attention be more premeditated.


At the training stage, the forward and backward positions are both derived from forced alignment labels. At the inference stage, the forward position is calculated according to the attention weights of preceding steps and the backward position is estimated from the predicted duration. To convert the forward \slash backward distance into a learnable vector, an embedding lookup layer is used in which two lookup tables are learned for forward and backward distances separately.

\subsection{Training Loss}
\label{ssec:loss}

The overall loss is a weighted sum of four parts as
\[ L = L_{mel} + \alpha _1 L_{pc} + \alpha _2 L_{dur} + \alpha _3 L_{align} \tag{2} \]
where \(L_{mel}\), \(L_{pc}\), \(L_{dur}\), and \(L_{align}\) denote MSE loss of Mel-spectrogram reconstruction, Cross-Entropy (CE) loss for the phoneme classifier, L1 loss for duration predictor, and MSE loss for guided attention, respectively. Their weights are set as \(\alpha _1=0.005, \alpha _2=0.025, \alpha _3=0.25\) empirically. 

Here, the stop token predictor \cite{Tacotron2} is not used. Instead, the decoder is assumed to stop when attention has stayed at the last token for the predicted duration time.

\section{Experiments}
\label{sec:expe}

\subsection{Training Setup}
\label{ssec:train}

We evaluated the proposed model on an internal corpus, which was from a non-professional female speaker, containing about 10 hours of speech data (about 12,000 utterances). The audios were collected in native mandarin Chinese and resampled into 16 kHz, 16-bit mono wave format.

A proprietary front-end engine was used to convert input texts into token sequences, which contain phonemes, tones, prosodic boundaries, and silence marks. Besides, a Kaldi-based forced alignment tool \cite{povey2011kaldi} was used to obtain phoneme duration labels from recordings.

Two variants of Tacotron2 are used as baselines. One replaces the attention mechanism in Tacotron2 with a duration informed length regulator (called TLR), and the other employs stepwise monotonic attention (called TSW). The postnet module is removed due to limited effectiveness. The reduction factor is set to 1 for a better quality of speech.

The same pre-trained LPCNet \cite{valin2019lpcnet} is used as a vocoder to generate audio signals from the predicted Mel-spectrogram.



\begin{table}[htb]
    \caption{The MOS with 95\% confidence intervals for the proposed method (PAMA), ground-truth samples (GT), and two baselines (TLR and TSW). The ground truth is obtained via analysis-synthesis.} 
    \centering
    \begin{tabular}{c c}
    \toprule
         Models & MOS \\
         \midrule
         GT & 4.54 \(\pm\) 0.12 \\
         \midrule
         TLR & 4.22 \(\pm\) 0.14 \\
         TSW & 4.38  \(\pm\) 0.18 \\
         \textbf{PAMA} & \textbf{4.41} \(\pm\) 0.14 \\
         \bottomrule
    \end{tabular}
    \label{tab:mos}
\end{table}



\begin{table}[htb]
\caption{The mean absolute errors (ms) of phoneme duration between the predicted by duration model and the segmented from the synthetic speech by forced alignment. The duration factor is used to scale the predicted duration to control the speech rate of synthetic speech.}
\centering
\begin{tabular}{c  c c c }
\toprule
\multirow{2}{3em}{Model} & \multicolumn{3}{c}{Duration Factor} \\
\cmidrule(l){2-4}
 & 0.75 & 1.0 & 1.5 \\
\midrule
TLR & 9.72 & 7.53 & 14.67  \\
TSW & - & 65.92 & -\\
\textbf{PAMA} & \textbf{8.48} & \textbf{6.68} & \textbf{13.54} \\
\bottomrule
\end{tabular}
\label{tab:mae}
\end{table}

\begin{figure}[htb]
  \centering
  \includegraphics[scale=1.0]{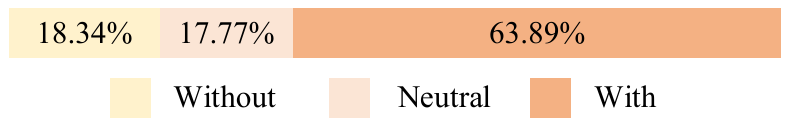}
  \caption{Preference test between PAMA-TTS with and without relative position embedding, which is at a \textit{p} \textless 0.01 level.}
  \label{fig:ab_test}
\end{figure}


    

\begin{table}[htb]
    \caption{The phoneme error rates (PER) of different models on 1,000 test sentences using a reduction factor (DF) of 1.0, 0.75 and 1.5. 
    For TSW, speed modulation is almost infeasible.}
    \centering
    \begin{tabular}{c c c c c}
    \toprule
    DF & Error & TLR & TSW & \textbf{PAMA}  \\
    \midrule
    
    \multirow{4}{2em}{1.0} & Sub & 1.92 & 1.68 & 1.69 \\
    & Del & 0.27 & 1.13 & 0.23 \\
    & Ins & 0.20 & 0.21 & 0.20 \\
    & PER & 2.39 & 3.02 & \textbf{2.12} \\
    \midrule
    
    \multirow{4}{2em}{0.75} & Sub & 4.23 & - & 3.16\\
    & Del & 1.84 & - & 0.93  \\
    & Ins & 0.37  & - & 0.42 \\
    & PER & 6.44 & - & \textbf{4.51}  \\
    \midrule

    \multirow{4}{2em}{1.5} & Sub & 1.60  & - & 1.43  \\
    & Del & 0.16 & - &  0.15  \\
    & Ins & 0.27 & - & 0.28  \\
    & PER & 2.03  &  - & \textbf{1.86}  \\
    \bottomrule
    \end{tabular}
    \label{tab:change_speed}
\end{table}

\subsection{Evaluation Setup}
\label{ssec:eval}

Two objective evaluations were conducted using 1,000 sentences. Firstly, the duration consistency was measured to show the duration controllability of models, which was calculated as the mean absolute errors (MAE) between the phoneme duration predicted by the duration predictor and that from a forced aligner. For TSW, phoneme duration was estimated from the attention results as \( d_i= \sum_{t=1}^{T} [arg max_n\alpha_{n,t}=i ]\), where \(d_i\) was the duration of the \(i\)th phoneme, and \(  \alpha \in \mathbb{R}^{N \times T} \) was the final attention matrix. Secondly, the phoneme error rate (PER) given by an automatic speech recognition (ASR) model was adopted as the metric to measure the robustness of different models. The ASR model was based on a TDNN-LSTM structure and trained on nearly 100,000 hours of recordings collected from various Xiaomi mobile phones.

Subjective evaluations were conducted using 30 sentences. They were not included in the training data. The naturalness of the synthetic speech was evaluated through the mean opinion score (MOS) test and AB preference test. 16 native listeners participated in the test, and the speech samples were shuffled in each test. 


\subsection{Results \& Discussion}
\label{ssec:result}

As shown in Table ~\ref{tab:mos}, the proposed model (PAMA) gets the highest mean opinion score. TLR shows slightly mechanical rhythm while TSW has clarity issues in some cases.

Results of the AB preference test shown in Fig. ~\ref{fig:ab_test} confirm the importance of procession-awareness for attention. If the relative position embedding is not leveraged, the naturalness of synthetic speech has remarkable degradation.


To check the duration controllability of different systems, MAE and PER are calculated for three duration factors (DF). We find stepwise monotonic attention is very weak at speech rate control. When attention score bias is shifted within a small range [-3, 3], the speech rate has a very slight change. However, if a greater shifting is applied, serious word skipping \slash repeating issues occur frequently. Therefore, only TLR and PAMA are evaluated for duration modification. Table ~\ref{tab:mae} compares the capability of duration control. It shows PAMA has on-par or even fewer duration errors than TLR, and an overwhelming advantage over TSW. Table ~\ref{tab:change_speed} compares the robustness with an ASR tool, in which PAMA has much fewer deletion errors than TSW and even better than TLR on overall performance.



\section{CONCLUSION}
\label{sec:conc}

This paper introduced progression-aware monotonic attention for robust sequence-to-sequence speech synthesis. The proposed model (PAMA-TTS) demonstrates that injecting the duration and relative position information into attention can achieve a better balance between the robustness and naturalness of synthetic speech. Besides, it enables accurate control of phoneme duration. Subjective and objective evaluation results show that PAMA-TTS outperforms the attention-based model on robustness and duration controllability while outperforms the duration-informed model on naturalness. Progression-aware monotonic attention is proved to be feasible for token length control and may be extended to other similar applications easily.  

\vfill\pagebreak

\bibliographystyle{IEEEbib}
\bibliography{strings,refs}

\end{document}